\documentclass[a4paper,12pt]{article}
\usepackage{graphics}

\newcommand{\abs}[1]{\left|{#1}\right|}
\newcommand{\refeq}[1]{Eq.~(\ref{#1})}

\title{Oscillation and levitation of balls\\
in continuously stratified fluids}

\author{I. Bir\'o%
\thanks{Faculty of Physics, Babe\c{s}--Bolyai University,
Kagolniceanu 1, 400084 Cluj-Napoca, Romania}%
,
B. Gy\"ure%
\thanks{von K\'arm\'an Laboratory for Environmental Flows, 
E\"otv\"os  University,  
H-1117, P\'azm\'any P. s. 1$/$A, Budapest, Hungary}%
, 
I. M. J\'anosi\footnotemark[2]%
,
K. G. Szab\'o%
\thanks{Department of Fluid Mechanics, Budapest University of Technology, 
H-1111, Bertalan L. u. 4--6, Budapest, Hungary}%
{$\;$}%
\thanks{%
Previous address: HAS Research Group for Theoretical Physics, 
H-1518, P.O. Box 32, Budapest, Hungary}%
, T. T\'el\footnotemark[2]%
}

\begin{document}

\maketitle

{\abstract{
The free motion of balls is investigated experimentally 
in continously stratified fluid
in a finite container. 
The oscillation frequency is found
to be very close to the local 
Brunt--V\"ais\"ala frequency. 
The effect of added mass proves to be practically negligible.
The evolution of rear jets is demonstrated, and a
kind of long term levitation is found. 
We show that the classical viscous drag would lead to a much 
stronger damping than observed in the experiment. This is interpreted 
as a consequence of the feedback from the previously excited internal waves 
following their reflection from the boundaries.  
A phenomenological equation with a modified drag term is proposed
to obtain a qualitative agreement with the observations. We 
point out 
that the inclusion of a history term would lead further away      
from the observed data. 

}}

\section{Introduction}

The motion of a spherical body in a fluid 
is one of the historical and still open problems of physics.
Its research dates back to the 18th century,
continuing with the works of Poisson, Green, Stokes 
and other outstanding scientists of the 19th century,
including the derivation of the famous history force
by Boussinesq \cite{Bous2} 
in 1885
(and, apparently independently, by Basset \cite{Bass1} 
three years later).
The correct mathematical formulation of these two nontrivial effects had been
the subject of research for about a century,
before the equation of motion,
valid at least for a small rigid sphere in an unbounded homogeneous fluid 
at low Reynolds numbers,
was finally settled 
by the work of Maxey and Riley \cite{MR}
(see also \cite{Max2}). 
Their approach has lead to a series of experimental 
and theoretical investigation of finite particle motion
in cases under more general conditions, as well
(for reviews see \cite{Mich, ME}, 
for a few recent examples, see 
\cite{Mordant, Falk} and \cite{Bec}, respectively).

An analogous problem is the motion of a rigid sphere in stably
stratified fluid.
In an early approach, before the publication of the Maxey--Riley equation,
Larsen \cite{Larsen} pointed out the importance of the generation of internal
waves and derived a linear equation for the oscillation of the sphere
about its stable equilibrium in an inviscid fluid 
when damping is entirely due to wave radiation.
More recently, the flow past a vertically moving ball has been studied
experimentally and numerically \cite{Ochoa, Torres}, 
and the appearance of a 'rear jet' was pointed out,
in which light fluid dragged down by the ball 
emerges from the boundary layer 
in a narrow column in the wake of the falling body.
Investigations of the gravitational settling of particles through sharp
density interfaces \cite{Srdic, Abaid} 
show that the light fluid surrounding the ball 
slows down the  settling, and might even lead to a 
temporary rising, 'levitation'
\cite{Abaid}, of the otherwise denser body.

A somewhat related approach is the study of internal wave generation
by vibrating obstacles \cite{Hurley,Erm,EG1,EG2}.  
For these forced types of motion the authors find in linear approximation
that both the added mass and the viscous drag 
are frequency dependent, 
and their actual value does also depend on
the ratio of the size of the moving body to the fluid depth.

In this paper we report upon our experimental investigation 
of the free motion of balls
in continously stratified fluid
in a finite container. 
The oscillation frequency of a fluid element is known to be given by the
Brunt--V\"ais\"ala (BV) frequency. The ball's frequency 
might, however, be different due to the added mass, history and other
effects. The basic questions raised in the paper are
\begin{itemize}
\item[-] is the oscillation frequency close to the BV frequency,
\item[-] is the damping primarily determined by the viscous drag,
\item[-] is an asymptotic state reached in the experiment,
\item[-] is there any equation which could faithfully describe the phenomenon?
\end{itemize}
We point our that 
in the course of the ball's motion, 
viscosity, wave generation 
and the reflection of internal waves from the boundaries 
all play an important role. 

In the next section different available forms of equations of motion are reviewed.
Then (Sections 3,4) we describe the experimental set-up and data
acquisition.  In Section 5 we present shadow-graphs showing the
evolution of rear jets during this nonmonotonic fall as well, and find a
kind of long term levitation. 
Section 6 is devoted to the analysis of the oscillation frequency which 
proves to be very close to the local BV frequency. This implies
that the effect of added mass is practically negligible during the 
late stage of the motion.
In Section 7 we show that classical viscous drag would lead to a much 
stronger damping than observed in the experiment. We interpret this 
as a consequence of the feedback from the previously excited internal waves 
following their reflection from the boundaries, 
and apply a phenomenological equation with a modified drag term
to obtain a qualitative agreement with the observations. The concluding 
section 
points out 
that the inclusion of a history term would lead further away      
from the observed data. A new term, whose explicit form remains unknown,
is, however, needed in a correct equation of motion, a term which 
accounts for the 
fluid motion generated by the internal waves.

\section{Theoretical background}

In an infinite homogeneous fluid of density $\varrho_f$,
the equation of motion of a small rigid spherical nonrotating 
particle of radius $r$ and mass $m_p$,
starting from rest at $t=0$,
is given by the Maxey--Riley equation ~\cite{MR,Max2}:
\begin{eqnarray}
  m_{p} \ddot{\mathbf{r}} & = & m_{f}    
\frac{D\mathbf{u}}{Dt}(\mathbf{r},t)
     +\frac{1}{2}m_{f} 
\left( \frac{D\mathbf{u}}{Dt}(\mathbf{r}, t) -  \ddot{\mathbf{r}} \right) 
+
(m_{p}-m_{f})\mathbf{g} \nonumber \\
 &  & - 6\pi r \nu \varrho_f 
\left( \dot \mathbf{r} -\mathbf{u}(\mathbf{r},t) \right)
+ \mathbf{F}_{\mathrm{history}}(t),
  \label{basic-equation}
\end{eqnarray}
where   
$\mathbf{r}(t)$
is the location of the particle at time $t$,
$\mathbf{u}(\mathbf{r},t)$ is the undisturbed velocity field of the fluid
in the absence of the ball 
(as determined by the boundary conditions and external sources),
$
{D\mathbf{u}}/{Dt}=
{\partial\mathbf{u}}/{\partial t}+
(\mathbf{u}\cdot{\nabla})\mathbf{u}
$
is the usual hydrodynamical time derivative of the velocity
following a fluid element,
$m_f$ is the mass of the fluid displaced by the sphere, and $\nu$ 
is the kinematic viscosity. 
The force terms on the right hand side represent
the hydrodynamical force, the added mass contribution,
the buoyancy corrected weight, the Stokes drag and the history term, respectively.
The equation is valid for small relative velocities,
when the history term
takes the form
\begin{equation}
\mathbf{F}_{\mathrm{history}}(t)=
 -6 r^{2}\varrho_f (\pi \nu)^{1/2} \int _{0}^{t}d\tau\,
  \frac	{\ddot{\mathbf{r}}(\tau)- 
		{d\mathbf{u}}(\mathbf{r}(\tau),\tau)/{d\tau}
	}{
		{(t-\tau )}^{1/2}
	}
  \label{int}
\end{equation}
where $
{d\mathbf{u}}/{dt}=
{\partial\mathbf{u}}/{\partial t}+
(\dot{\mathbf{r}}\cdot{\nabla})\mathbf{u}
$
is the time derivative following the path of the particle.
This term is due to the
fact that the particle modifies the flow locally.

The original equation of Maxey and Riley~\cite{MR,Max2} contain further correction 
terms for strongly nonuniform background flows.
These terms have been omitted from \refeq{basic-equation} for simplicity,
since in what follows we consider only the case when the unperturbed 
fluid is at rest: $\mathbf{u}(\mathbf{r},t)\equiv 0$.
We shall also assume that the motion takes place 
in the vertical direction, along the $z$ axis.

At large particle velocities the Stokes drag is replaced
by the nonlinear drag
\begin{equation}
 - c_D(Re) \frac{r^2 \pi}{2} \varrho_f
  \abs{\dot{z}}\dot{z} 
 \label{drag}
\end{equation}
where $c_D$ is the empirically known drag coefficient \cite{drag},
a function
of the instantaneous Reynolds number  
\begin{equation}
Re=\frac{2 r \abs{\dot{z}}}{\nu}. 
\label{Re}
\end{equation}
In this regime, the form of the history term is known to differ from
(\ref{int}), and it is doubtful that it can be expressed at all
as a convolution
with a simple kernel. In any case it appears to decay faster then in the
Stokes regime (see e.g. \cite{Mordant}).  


We extend Eqs.\ (\ref{basic-equation}) and (\ref{int}) 
to allow spatially varying fluid densities
$\varrho_f(z)$ 
as follows
\begin{equation}
 \ddot{z} = - \sigma \frac{\varrho_f(z)}{\varrho_p} \ddot{z}  -
\frac{\varrho_p-\varrho_f(z)}{\varrho_p} g  
 -  c_D(2 r \abs{\dot{z}}/\nu)
    \frac{3 \varrho_f(z)}{8 r \varrho_p}\abs{\dot{z}}\dot{z} \,, 
  \label{Beq}
\end{equation}    
where $m_p = 4\pi\varrho_p r^3/3$ has been used with
$\varrho_p$ 
as the particle density.
Since the coefficient of
the added mass effect is not a unique constant in stratified flows, 
we replaced $1/2$ by an unknown coefficient $\sigma$. 
The history term has not been written
out since its contribution is found to be negligible in our experiment.
We shall return to a discussion of the relevance of this term 
in the Conclusions.       

According to \refeq{Beq} the particle is in a stable equilibrium at the height
where $\varrho_f(z)=\varrho_p$, which we choose as a convenient 
reference height $z=0$ in the following.
Assuming small amplitude oscillation around this neutrally buoyant position, 
we can write 
$\varrho_p-\varrho_f(z)=
z \varrho_p N^2/g$, where 
$N$ denotes the local Brunt--V\"ais\"ala (BV) frequency. 
If the amplitude is small enough, the velocity amplitude can also be
small enough to ensure the Stokesian regime all the time,
the drag coefficient is then $c_D(Re)=24/Re$
\cite{Kundu}. 
Thus we obtain
\begin{equation}
(1+\sigma) \ddot{z} =  
-N^2  z - 
2 \alpha_0 \dot{z},
  \label{Leq}
\end{equation}    
where 
\begin{equation}
\alpha_0=9 \nu/(4 r^2)
\label{alpha}
\end{equation} 
is a damping coefficient. 
The motion is then that of a linearly damped harmonic oscillator
with frequency $\omega_0$, given by
\begin{equation}
\omega_0^2 =  
\frac{N^2}{1+\sigma} - \frac{\alpha_0^2}{(1+\sigma)^2}.
  \label{omega}
\end{equation}     
It is remarkable that both viscosity and added mass effect tend to
decrease the oscillation frequency below the BV value. 
An important dimensionless parameter of the problem is 
\begin{equation}
St = \frac{\nu}{r^2 N}=\frac{2 Fr}{Re},   
  \label{St}
\end{equation}    
which can be considered as a Stokes number. 
$St$ is basically the ratio
of the instantaneous Froude number, 
\begin{equation}
Fr=\frac{\dot{z}}{N r}
\label{Fr}
\end{equation}  
and the Reynolds number. 

An alternative approach is due to Larsen \cite{Larsen}
who derived for a small amplitude oscillations of a
sphere, released initially at height $z(0)$: 
\begin{equation}
\ddot{z} =  
-N^2  z +
  \frac{z-z(0)}{t^2} -\frac{\dot{z}}{t}.
  \label{Larsen}
\end{equation}    
The last two terms on the right hand side express energy loss due to
radiation of internal waves, an effect not taken into account
in (\ref{Beq}). This approach neglects, however, viscous effects.
Due to the presence of the inhomogeneous terms, no clear 
oscillation frequency can be defined for short times. 

At present, no equation is known which would be able to account for 
nonlinear internal wave generation during the ball's motion.

\section{Experimental setup and density profiles}

The experiments were carried out in a glass tank of size
$75\;\mathrm{cm}\times 38\;\mathrm{cm}\times50\;\mathrm{cm}$. 
The salt density stratification
was produced by a double-bucket equipment \cite{Fort}. 
A typical water height of 38 -- 39~cm was used. 

The ambient density profile $\varrho_f(z)$ was obtained
by measuring the conductivity and the temperature of the salt solution at different
heights $z$, and by extracting the density values from tabulated data.     

The average BV 
frequency, $\bar{N}$, 
was deduced from the average slope of
the density profile. Table \ref{T:1} indicates these frequencies in the five
different profiles used.

\begin{table}[h!]
\centering
\begin{tabular}{|c|c|}
  \hline
  Profile & $\overline{N}$ (1/s) \\  \hline
  1& $1.21\pm 0.02$ \\
  2& $1.23\pm 0.02$ \\
  3& $0.86\pm 0.01$ \\
  4& $0.58\pm 0.01$ \\ 
  5& $1.12\pm 0.03$ \\ \hline
\end{tabular}
\caption{The average BV frequency for the different density profiles used.} 
\label{T:1}
\end{table}

\begin{figure}[h!]
{\hfill
\resizebox{70mm}{!}
{\includegraphics{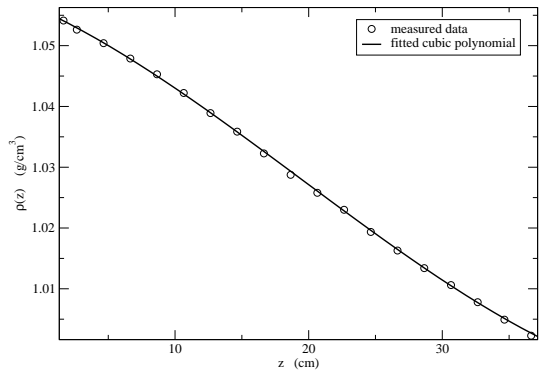}}
\hfill}
  \caption{The density profile no. 1 (cf. Table \ref{T:1}).
The cubic polynomial fitted to the data is
$\varrho_f(z)=1.0559-1.02 \times 10^{-3} z-
3.14 \times 10^{-5} z^2 + 5.39 \times 10^{-7} z^3$,
where density and height are measured in g/cm$^3$ and cm, respectively.
}
  \label{F:1}
\end{figure}

A careful investigation of the profiles indicated that there was a slight
static deviation superimposed on the linear slope. 
In order to obtain a more precise
expression for the local BV frequency, $N(z)$, at different heights, we
fitted a cubic polynomial to the density data, which proved to be an 
appropriate form for all the profiles.
A typical profile and the fitted polynomial is shown in Fig. \ref{F:1}.

The motion of five different balls of radius $r=7.3\;\textrm{mm}$ was followed
in the tank.
The plastic balls were prepared by implanting small metal pieces
right below their surface.
This arrangement,
with substantial distance between the center of gravity 
and the geometrical center of the ball,
yielded a strong uprighting tendency of the submerged balls,
and helped stabilizing its attitude and avoiding rotation.
The density $\varrho_p$
of the balls (see Table \ref{T:2}) was adjusted so
that the balls had a neutral position within the tank 
for most density profiles.
With a given ball up to $4$ 
experiments have been carried out in the same tank subsequently.

\begin{table}[h!]
\centering
\begin{tabular}{|c|c|}
  \hline
  Ball & $\varrho_p$ (g/cm$^3$) \\  \hline
  1& $1.009(9)$ \\
  2& $1.016(5)$ \\
  3& $1.025(8)$ \\
  4& $1.038(0)$ \\ 
  5& $1.047(2)$ \\ \hline
\end{tabular}
\caption{The density of the balls used.} 
\label{T:2}
\end{table}

The balls were initially kept fixed at the end of a tube 
connected to a vacuum pump,
slightly below the water surface. The motion 
was initiated by gradually diminishing the vacuum. 
We evaluated those ball paths only which exhibited a very weak drift in the
horizontal direction. 
In some cases, however, a strong drift evolved
right after the initiation of the motion 
as a feedback of the nonlinear vortices and lee waves generated.

\section{Data acquisition}

\begin{figure}[h!]
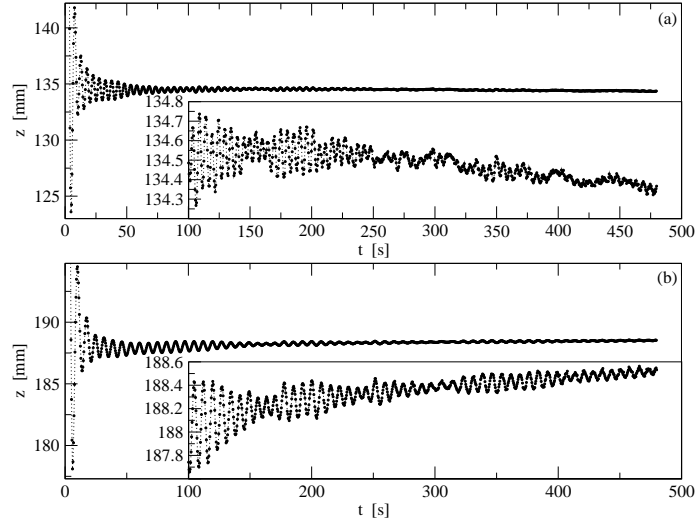

{\hfill
\resizebox{90mm}{!}{\includegraphics{3a_abra.eps}}
\hfill}

{\hfill
\resizebox{90mm}{!}{\includegraphics{3b_abra.eps}}
\hfill}
  \caption{Height vs. time diagrams. To guide the eye, 
measured points have been joined by dotted lines.
The insets show the curves with higher vertical resolution
after $100$ seconds.
(a) Profile 2, Ball 4, (b)Profile 3, Ball 2. 
}
  \label{F:2}
\end{figure}

The dynamics was monitored by digital cameras (Sony DCR-PC 115E PAL, 
PCO Pixelfly).
The location of the ball was determined 
as the 'center of mass' of the black pixels 
representing the ball on a digitalized image. 
With this method the vertical position of the body
could be determined with a resolution of approximately 
0.1 mm. 
Two typical height vs.\ time 
plots obtained this way are shown in Fig. \ref{F:2}.

The figure clearly indicates a strong damping of the ball
in the course of the first 3 -- 4 oscillations. 
The intermediate and long time behavior is, however, not a simple 
damped oscillation since, as the insets indicate, the motion appears to be a 
superposition of several oscillations. It is remarkable that a really steady position
has not been reached even after more than 500 seconds.   
The insets in Fig. \ref{F:2} also show a gradual slow change of the average height, either upward or downward. We discuss this point in more detail
in Section 5.

\begin{figure}[h!]
{\begin{center}
\resizebox{90mm}{!}{\includegraphics{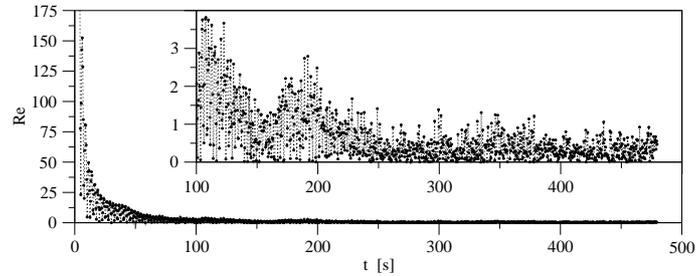}}
\end{center}}
  \caption{Reynolds number vs. time diagram for the
experiment of Fig. \ref{F:2}a. To guide the eye, 
measured points have been joined by dotted lines.
The inset shows the curve with higher vertical resolution
after $100$ seconds.
}
  \label{F:3}
\end{figure}

In order to get insight into the importance of viscous effects 
during the motion, we show in Fig. \ref{F:3} the Reynolds number 
vs. time of one of these sample experiments. 
The plot clearly indicates that the Reynolds number 
(\ref{Re}) 
reaches the value unity at around
200 seconds, and it never drops really much below this value. 
It appears that Stokes regime is hardly ever reached. 
The Froude number 
(\ref{Fr}) is $\nu/(2{N}r^2)=St/2$
times larger than Re, and since this ratio is approximately $10^{-2}$ with our data, 
we conclude that the Froude number is much below unity during the entire motion,
which indicates the importance of the stratification effects. 

\begin{figure}[h!]
{\hfill
\resizebox{90mm}{!}{\includegraphics{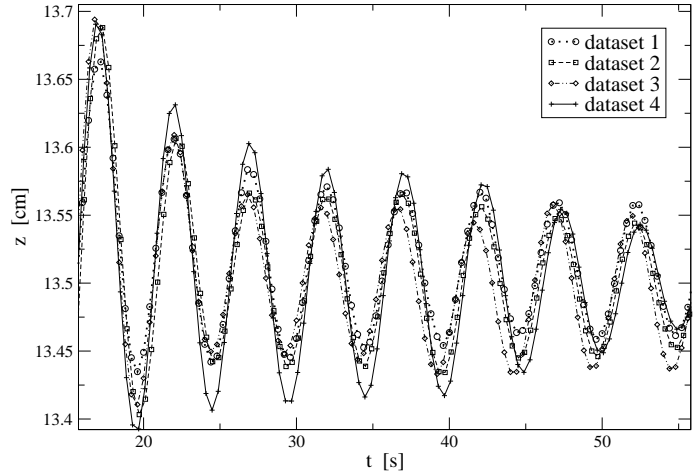}}
\hfill}
  \caption{Height vs. time diagrams
with Ball 4 in Tank 2 for 4 initial conditions as identical as possible
with our experimental accuracy.
The time interval shown starts with the third oscillation of the ball.
The measured dots are connected with different types of lines.
}
  \label{F:4}
\end{figure}

It needs to be mentioned that the experiments are not
reproducible in detail. To illustrate this, Fig. \ref{F:4}
shows the results of four experiments repeated 
with identical initial conditions
and other parameters. The curves 
are shifted in order to reach a collapse
as good as possible. 
The amplitudes are markedly different, but the frequencies 
are the same with a reliable accuracy. 
In the next Section we will show that the observed 
irreproducibility is a consequence
of irregular lee vortex and wave generation, and not of an 
instrumental artefact. 
The main evaluated quantity of our measurements
will therefore 
be the frequency of the oscillations.  
  
\section{Shadow-graphs and levitation}

By illuminating the balls with a strong directed
light beam, a clean shadow-graph 
picture of the flow pattern appears on a plane paper adjusted to the back
wall of the container. 

\begin{figure}[h!]
\resizebox{0.95\textwidth}{!}{\includegraphics{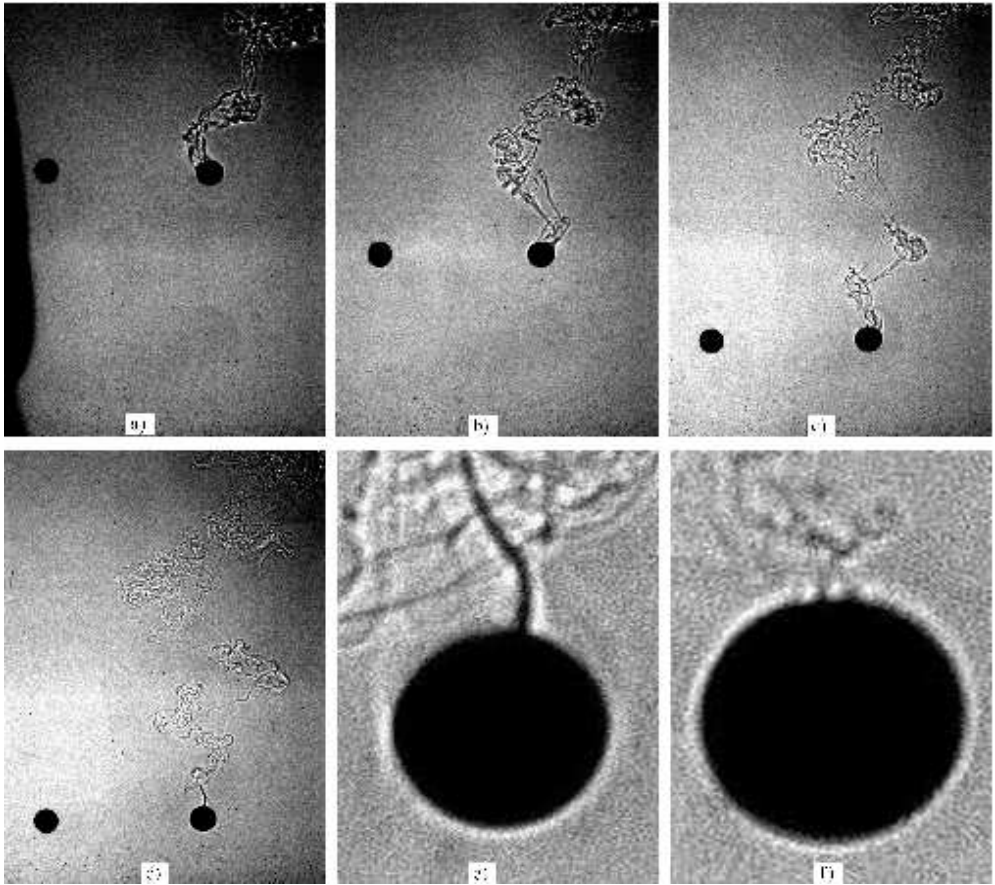}}
  \caption{
Shadow-graph images over the first period of motion of ball 3 in Tank 3.
The time spent after release is:
a) $4$ s, b) $6.4$ s, c) $10$ s, d) $16$ s, e) $22$ s, f) $54$ s. 
The direct image of the ball of diameter $14.6$ mm
appears on the left of the shadow.
}
  \label{F:5}
\end{figure}

In the first part of the motion, the rapid falling of the balls,  
strong density inhomogeneities appear, indicating
an irregular generation of lee waves
(Figs \ref{F:5}). 
The characteristic wavelength 
is on the order of 5 cm in both the horizontal and the vertical
direction.  
It is this strongly nonlinear set of events which can lead to the gain of a 
horizontal momentum of the ball, mentioned previously. 
The apparent random nature of the lee vortex and wave generation explains 
also the amplitude anomaly shown in Fig.\ \ref{F:4}.
The details of the patterns in subsequent runs are very different,
consequently, vertical damping and the structure of the
dragged boundary layer can also be very different.
These phenomena, although different in several details, 
can be paralleled
with the chain of events occurring
behind spheres in homogenous flows at similar Reynolds numbers.


During the second part of the motion,
shown on panels d)-f) of Fig.\ \ref{F:5}, 
when the oscillation amplitude is already small,
a couple of unparallel consequences of the stratification can be identified.

First, a boundary layer
around the ball 
consisting of light fluid 
dragged from the upper fluid regions
in the course of the fall is clearly distinguishable.
Second,
the pictures also show the 
narrow rear jet meandering upwards from the top of the sphere.
The jet ejects 
the light fluid 'evaporating'
from the surface of the ball back to the higher layers of the fluid.
Note how the intensity of the jet diminishes  with time
as its light fluid supply in the boundary layer decreases.
The slow 'evaporation' of this layer explains the long
term sinking of the balls (see e.g. Fig. \ref{F:2}a).  
These phenomena matches qualitatively to those found numerically
in \cite{Torres}.

In some other cases, however, we rather found 
a slow asymptotic rise of the ball. 
One possibility for this could be the gradual swelling
of the material of the ball due to its constant exposure to the liquid,
such an event has been reported in \cite{JST}.
In our case, the balls do not absorb water, thus
their slow rising were attributed to the attachment of tiny gas bubbles
to the surface of the ball. 
This effect is also random-like, unpredictable.
It was impossible to avoid or to control it
by using deaerated water since air
becomes dissolved in the fluid during the process of filling up
the tanks.

\begin{figure}[h!]
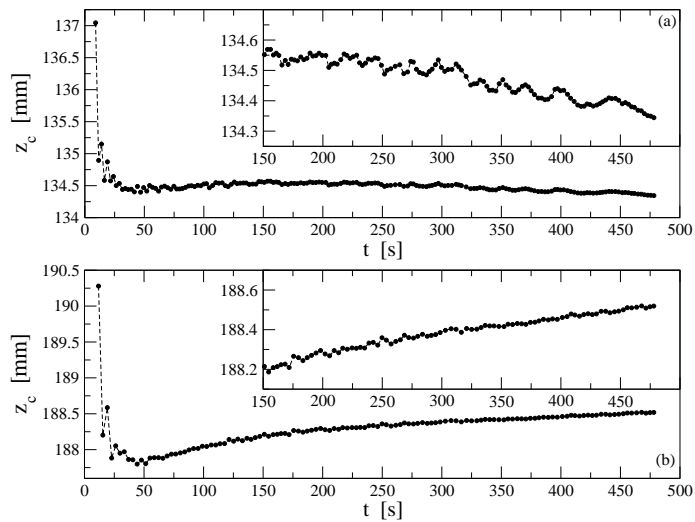

{\hfill
\resizebox{90mm}{!}{\includegraphics{8a_abra.eps}}
\hfill}

{\hfill
\resizebox{90mm}{!}{\includegraphics{8b_abra.eps}}
\hfill}
  \caption{
Time evolution of the centres of oscillation $z_c$
for the motion displayed in (a) Fig. \ref{F:2}a and (b) Fig. \ref{F:2}b.
}
  \label{F:6}
\end{figure}

For a more detailed study of the rising or sinking motion, we evaluated 
the centres of oscillation $z_c$ over half periods, i.e, the mean values of subsequent
displacement extrema. As Fig. \ref{F:6} indicates, the oscillation centers
always sink drastically over the first $5-6$ periods. 
This is due to the active 'evaporation' of the light fluid from the surface, 
which continuously increases the effective density of the ball. 
Then a deepest point is reached due to an oscillation overshoot
which is always 
followed by a slow rising, levitation, 
between about 50 and 200 seconds. 
At this stage there is still a
very narrow layer of light fluid around the ball, and hence the ball 
might rise beyond 
the height defined by the condition 
$\varrho_p=\varrho_f$.
This is followed by a sinking (Fig. \ref{F:6}a) if
the bubble formation is weak, otherwise, a continuous rising follows
(Fig. \ref{F:6}b).

\section{Oscillation frequencies}

\begin{figure}[h!]
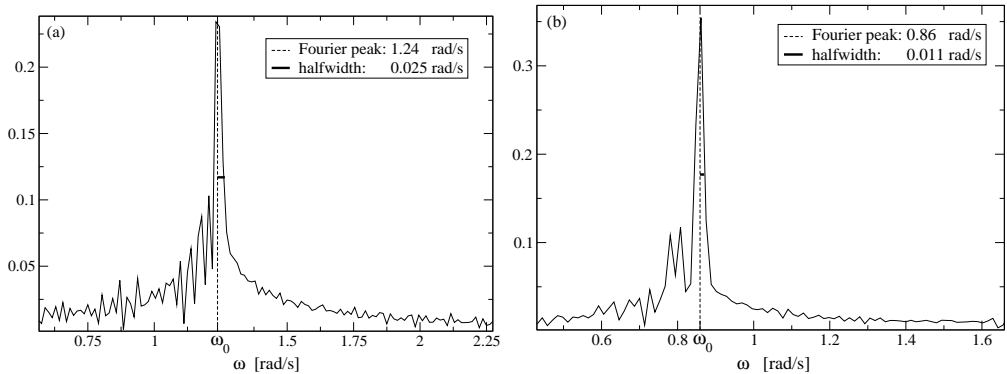

\resizebox{65mm}{!}{\includegraphics{10a_abra.eps}}
\resizebox{65mm}{!}{\includegraphics{10b_abra.eps}}
  \caption{
The power spectrum of the signals
of (a) Fig. \ref{F:2}a and (b) Fig. \ref{F:2}b. The halfwidths 
indicating the error are marked by short horizontal intervals.
}
  \label{F:7}
\end{figure}

The typical frequency $\omega_0$ of the oscillation was determined both by measuring the 
average period between local maxima of the displacement and by locating
the peak of the power spectrum of the function $z(t)$.
Fig. \ref{F:7} shows the Fourier peaks. The error estimate of the
frequency is given by the halfwidth of the peak.

\begin{figure}[h!]
{\hfill
\resizebox{100mm}{!}{\includegraphics{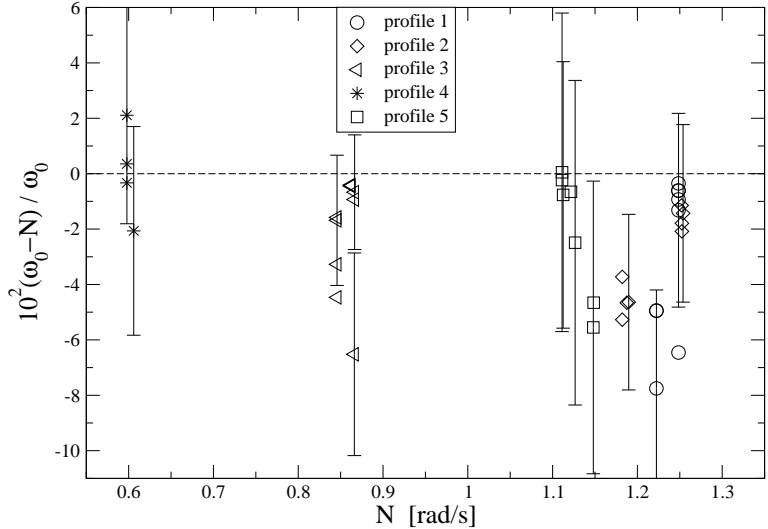}}
\hfill}
  \caption{
The relative deviation $(\omega_0-N)/\omega_0$
of the measured oscillation frequency from the local BV frequency vs the local BV 
frequency $N$.
Different symbols mark different profiles
(cf. Table \ref{T:1}). The average relative deviation is 
2 percents, indicating
an added mass coefficient $\sigma = 0.04$ at most. In individual cases
the $\sigma$ value turned out the be very close to zero. 
In order to avoid scatter,
only two representative error bars per profiles are shown.
}
  \label{F:8}
\end{figure}

Both methods mentioned above lead to practically identical oscillation
frequencies for a given measurement. The $\omega_0$ values fall rather close
to the local BV frequency, $N$, of the height around which the long term
oscillation took place. In order to see more detail, in Fig. \ref{F:8}
we plot the relative deviation $(\omega_0-N)/\omega_0$ from the BV frequency vs.
the local BV frequency itself. There is no clear trend visible 
on the plot, and the average deviation is on the order of 2 percents,
indicating that the added mass effect is rather weak. 
Note that for small added mass coefficient $\sigma$ and negligible damping coefficient
(with our data $\alpha_0=0.042$ 1$/$s), relation (\ref{omega}) yields 
\begin{equation}
\omega_0=\frac{N}{{(1+\sigma)}^{1/2}} \approx N(1-\sigma/2).
\end{equation}
The added mass coefficient 
in our experiments
thus can be estimated to be $\sigma \approx 0.04$.  
In view of the magnitude of the error, we can say that
the added mass effect is practically negligible.  
This observation seems to be consistent with
those of Ermanyuk and Gavrilov \cite{EG2} who
found in a linear setting that the added mass coefficient around the BV 
frequency tends
to zero with the increasing depth of the fluid.

\section{Amplitude dynamics}

\begin{figure}[h!]
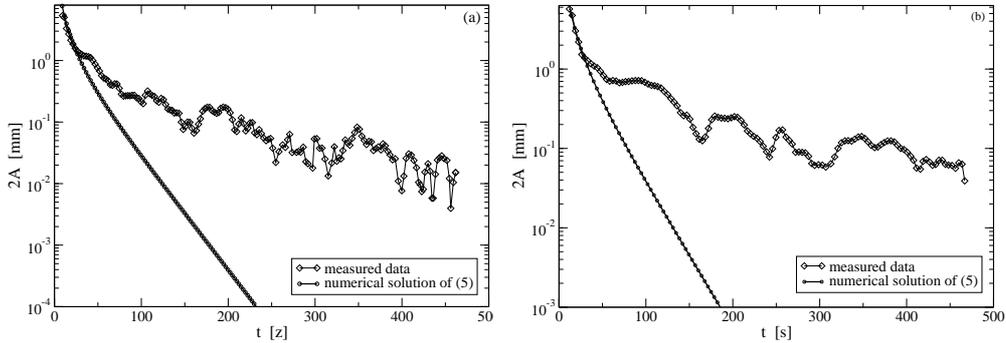

\resizebox{65mm}{!}{\includegraphics{12a_abra.eps}}
\resizebox{65mm}{!}{\includegraphics{12b_abra.eps}}
  \caption{
Local displacement maxima $2A$ measured as
the difference between consequtive maxima and minima of $z$ as
a function of time $t$.
Dots represent measured data, while the continuous curve is the
numerical solution of (\ref{Beq}) with the first 
local maximum of the measured height
as the initial location and zero initial velocity.
Panels (a) and (b) correspond 
to the measurement of Fig. \ref{F:2}a and b,
respectively.    
}
  \label{F:9}
\end{figure}

Although, as indicated earlier, the amplitudes of oscillations
do not reproduce properly, these quantities can be used to obtain 
global information about the ball dynamics. 
In Fig. \ref{F:9} we compare measured data with the numerical simulation
of equation (\ref{Beq}), where the added mass effect
is assumed to be negligible: $\sigma=0$. 
In order to have a clear comparison with the
exponential long term damping predicted by (\ref{Leq}), we plot the 
local displacement maxima on a logarithmic scale. The initial decay appears 
to be similar, but
after about a few tens of seconds a drastic 
deviation occurs: the measured data do not
follow the exponential rule, they exhibit a much weaker decay.
The ratio of the measured to computed 
amplitudes is on the order of $10^4$ at $t \approx 400$ s.

This strong deviation is attributed to the generation of lee waves by the
balls, in particular during the first period of their fall. These waves 
are reflected back from
the boundaries and interact with the ball upon their return.
Approximating the wave speed with that of linear internal waves, given by
\cite{Kundu}
\begin{equation}
c= \frac{N}{(k^2+m^2)^{1/2}},
\end{equation}
with $k$ and $m$ as the horizontal and the vertical wave numbers, respectively,
we obtain with $N \approx 1$ 1$/$s and $k=m \approx 2 \pi/(5$cm) 
a wave speed of $c\approx 6$ mm$/$s.
The time needed to return from a wall at a distance of a few dm is then 
on the order of $100$ s. This is the time when 
interference
patterns first appear in Fig. \ref{F:2}.
The viscous damping time of these waves 
is $(\nu(k^2+m^2))^{-1}$ which is on the order of $30$ seconds,
but the damping time for waves of dm wavelength
goes beyond $100$ seconds.  
This should be 
contrasted with the Stokesian damping time
$1/\alpha_0$  of (\ref{Beq}), which is $24$ seconds.
This is the time when the deviation
from the numerical solution sets in
in Fig. \ref{F:9}.

The presence of internal waves was experimentally verified
in a separate qualitative control experiment, by 
illuminating fluorescent dye layers in the fluid.
%

We conclude that the fluid motion is not negligible even 
long times after the initiation of the ball's motion, due to the presence
of excited and reflected internal waves. The assumption 
$\mathbf{u} \equiv 0$, used in the derivation of (\ref{Beq}),
therefore does not hold. After some time, the agitation by internal waves
overcomes viscous damping and makes the ball oscillating much stronger than
in a fluid at rest. 
This slow oscillation also takes place with approximately the 
BV frequency. 

\begin{figure}[h!]
{\hfill
\resizebox{90mm}{!}{\includegraphics{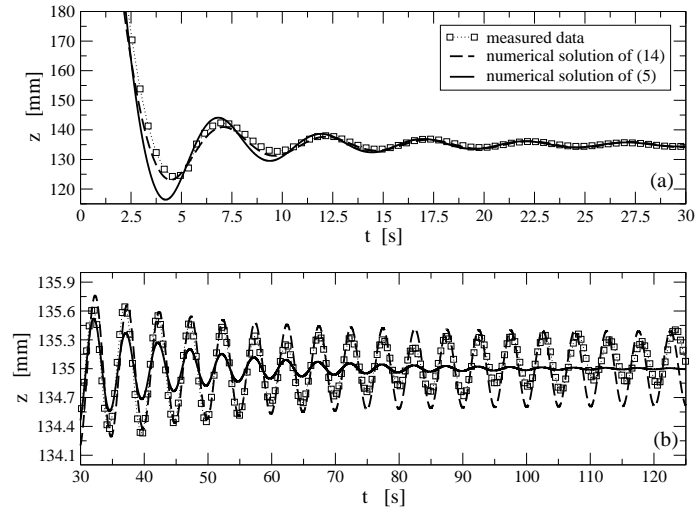}}
\hfill}
  \caption{
Comparison of a measured displacement vs. time 
(dataset (2) of Fig \ref{F:4}) with 
two simulations. Simulation 1 is carried out with (\ref{Beq}), while
Simulation 2 corresponds to (\ref{Beqm}) with $A=1.6$ and $\alpha_0=0.042 \ 1/s$. 
In both cases $\sigma=0$. 
Panel (a) and (b) correspond to the early and late stage of oscillations, respectively.
}
  \label{F:11}
\end{figure}

A phenomenological way of taking this into account in Eq. 
(\ref{Beq})
is the switching off of the viscous damping, by replacing $c_D$ 
proportionally to 
$c_D(Re) e^{(-\alpha_0 t)}$, 
where $\alpha_0$ is the damping coefficient (\ref{alpha}).
By solving equation  
\begin{equation}
 \ddot{z} = - \sigma \frac{\varrho_f(z)}{\varrho_p} \ddot{z}  -
\frac{\varrho_p-\varrho_f(z)}{\varrho_p}g  
 - A e^{-\alpha_0 t} \frac{3 \varrho_f(z)}{8 r \varrho_p}
\mid \dot{z} \; \mid \dot{z} c_D(2 r \mid \dot{z} \mid/\nu),
  \label{Beqm}
\end{equation}    
a reasonable agreement with the observed data is obtained, even for long times
(see Fig. \ref{F:11}). 
Coefficient $A$ should be chosen to obtain the best fit with the data.
We do not claim 
that this phenomenological equation would provide the best model for 
the ball's motion. The procedure, however, illustrates that
the naive equations should be modified in view of the long lasting
presence of internal waves.  

\begin{figure}[h!]
{\hfill
\resizebox{90mm}{!}{\includegraphics{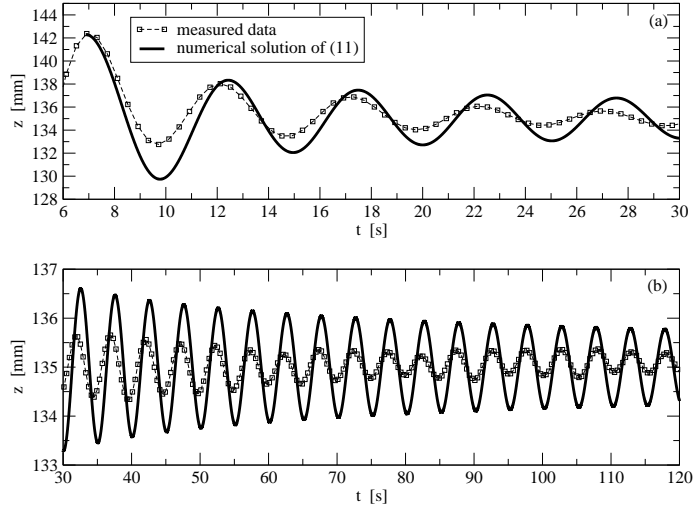}}
\hfill}
  \caption{
Comparison of measured displacement vs. time 
(dataset (2) of Fig \ref{F:4})
with 
Larsen's theory (continuous line), the simulation of 
(\ref{Larsen}). 
Panel (a) and (b) correspond to the early and late stage of oscillation
s, respectively.
}
  \label{F:11a}
\end{figure}

To conclude this Section, we present the results of
solving Larsen's equation (\ref{Larsen}). Fig. \ref{F:11a} shows the 
$z(t)$ function obtained by starting the simulation at the first maximum upward
displacement of the measured paths. In contrast to Larsen's original
observations \cite{Larsen}, 
valid for very small amplitude oscillations, the numerical
solution markedly deviates from the data in the course of
the first oscillations already. The frequency appears to be smaller than
the measured one. In addition, the long term behavior is much more damped than 
the numerical solution.

\section{Discussion and Conclusion}

\begin{figure}[h!]
{\hfill
\resizebox{90mm}{!}{\includegraphics{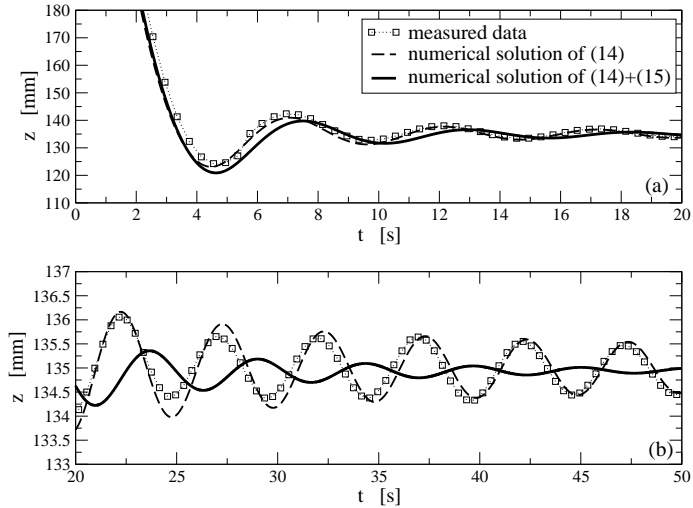}}
\hfill}
  \caption{
Comparison of the measured displacement vs. time data 
of Fig \ref{F:2}a
with 
simulations with and without the history term. 
Simulation 1 is carried out with (\ref{Beqm}), 
and Simulation 2 is the same with
term (\ref{hist}) added to the right hand side. 
In both cases $\sigma=0$, $A=1.6$ and $\alpha_0=0.042 \ 1/s$. 
Panel (a) and (b) correspond to the early and late stage of oscillation
s, respectively.
}
  \label{F:12}
\end{figure}

We stated that the history term (\ref{int}) is unimportant 
in (\ref{basic-equation}) in our experiment. One reason for this is that the 
relative acceleration which appears in the integrand 
is small in the long time dynamics, at least.
Here we show 
that the presence of the history term
does not lead to a better agreement with the measured behavior for
short times either. Ignoring the doubts that (\ref{int})
needs not be valid for Reynolds numbers much larger than unity,
we solved Eq.\ (\ref{Beqm}) with the term
\begin{equation}
- \frac{9}{2r}\sqrt{\frac{\nu}{\pi \varrho^2_p}}
\int_0^t
\varrho_f(z(\tau)) \frac{\ddot{z}}{(t-\tau)^{1/2}} \; d\tau
\label{hist}
\end{equation}
added numerically. 
The simulation clearly indicates (Fig. \ref{F:12})
a much stronger deviation from the measured data than without
this term: the frequency is smaller then
the measured one, and damping is much stronger.

\begin{figure}[h!]
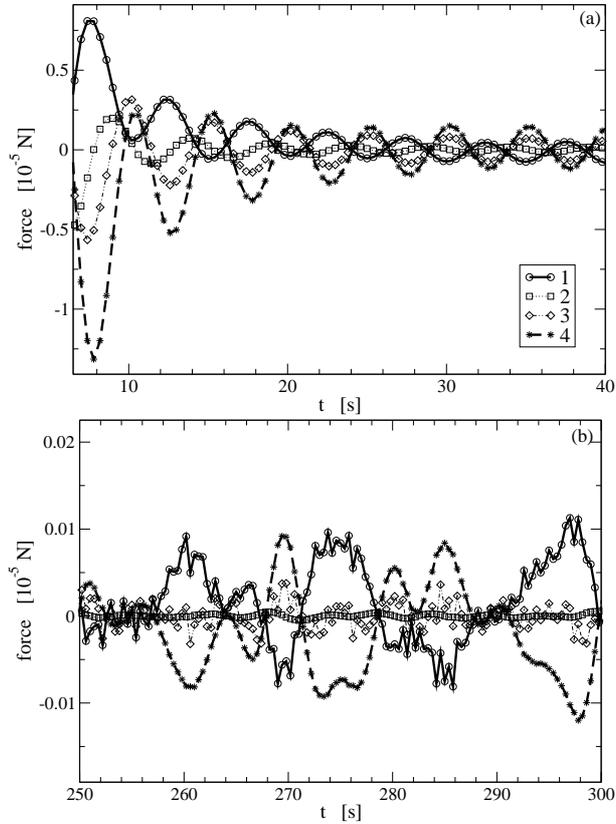

{\hfill
\resizebox{80mm}{!}{\includegraphics{16a_abra.eps}}
\hfill}

{\hfill
\resizebox{80mm}{!}{\includegraphics{16b_abra.eps}}
\hfill}

  \caption{
Contribution of the rest force 
and of different terms in (\ref{rest}) in the time
series of Fig. \ref{F:2}a. 1: rest force $F_r$; 
2: negative drag $(ii)$; 
3: mass times acceleration $(i)$; 
4: buoyancy corrected weight $(iii)$.  Added mass is neglected.
Panel (a) and (b) show the first $40$ s and the 
late stage after $250$ s, respectively.
}
  \label{F:13}
\end{figure}

The fact that the history term is not appropriate for interpreting observed 
data does not imply of course that Equation (\ref{Beq}) is valid.
To obtain a measure of the lack of equality 
multiply \refeq{Beq} by $m_p$ and rearrange all terms to the left hand side
to get
\begin{equation}
F_r = \underbrace{m_p \ddot{z}}_{(i)}
+ \underbrace{{\pi r^{2}}
c_D(2 r \abs{\dot{z}}/\nu)
\varrho_f \abs{\dot{z}} \dot{z}/2}_{(ii)} 
+ \underbrace{(m_{p}-m_{f}(z)){g}}_{(iii)}. 
\label{rest}
\end{equation}
We applied numerical differentiation to the measured data and
evaluated each term $(i)$-$(iii)$ in (\ref{rest}) separately,
thus this formula provides us with a `rest force' $F_r$. 
The result is exhibited in Fig.\ \ref{F:13}.

The result clearly indicates that the rest force is never negligible. For long times the
drag and the acceleration become very small, and hence
the rest force is practically the negative of the buoyancy corrected weight. 
This further supports our view that the long 
time motion of the ball is very close to a passive advection in the fluid, 
with a very small velocity difference $\dot{\mathbf{r}}(t)-\mathbf{u}({\mathbf{r}}(t),t)$.
In view of (\ref{basic-equation}), it is the
term $\varrho_f D \mathbf{u}/D t$, i.e., 
the acceleration of the ambient fluid in the container, 
which is missing from (\ref{Beq}).  

Note that by dropping the naive assumption $\mathbf{u} \equiv 0$,
the paradox of non Stokesian asymptotical behaviour mentioned in Section 4
can also be solved. If the background velocity field is nonzero,
the definition (\ref{Re}) for 
the Reynolds number shall be 
replaced by the proper one which is based on the relative velocity 
rather than on $\abs{\dot{z}}$.
In the limit of passive advection this Reynolds number
approaches 0, i.e. the ball enters  Stokesian regime asymptotically.

The ambient velocity $\mathbf{u}$ appears due to the internal
waves excited by the falling ball, but it also has a feedback on the ball.
The problem of the velocity field
and of the particle velocity cannot be decoupled,  they should be treated
self-consistently. In stratified fluids, therefore, there is no hope for
any simple equation for ${\bf r}(t)$ which would contain the particle-free
fluid velocity.

\section{Acknowledgements}

The authors are indebted for Detlef Lohse for a fruitful discussion.
This work was supported by the Hungarian Science Foundation
(OTKA) under grants TS044839 and T047233. IMJ thanks for a
J\'anos Bolyai research scholarship of the Hungarian
Academy of Sciences.



\end{document}